\documentclass[aps,prd,jmp,reprint,twocolumn,superscriptaddress]{revtex4}
\usepackage{hyperref}
\usepackage{footnote}
\usepackage{amssymb}
\usepackage{amsmath}
\usepackage{graphicx}
\usepackage{dcolumn}
\usepackage{bm}
\usepackage{color}
\usepackage{xcolor}
\usepackage{float}
\usepackage{lipsum}
\allowdisplaybreaks
\usepackage{subfig}
\usepackage{comment}
\usepackage{multirow}
\usepackage[figurename=FIGURE,tablename=TABLE]{caption}
\usepackage{threeparttable}
\usepackage{hyperref}
\hypersetup{
    colorlinks=true, 
    linktoc=all,     
    linkcolor=magenta,
    citecolor=red
}

\newcommand{\ba}{\begin{eqnarray}}
\newcommand{\ea}{\end{eqnarray}}

\begin{document}

\title{Traversable Wormholes in Einstein-Euler-Heisenberg Gravity: Geometry, Energy Conditions, and Gravitational Lensing}

\author{Phongpichit Channuie}
\email{phongpichit.ch@mail.wu.ac.th}
\affiliation{School of Science, Walailak University, \\Nakhon Si Thammarat, 80160, Thailand}
\affiliation{College of Graduate Studies, Walailak University, \\Nakhon Si Thammarat, 80160, Thailand}

\author{Allah Ditta}
\email{mradshahid01@gmail.com}
\affiliation{Department of Mathematics, School of Science, \\University of Management and Technology,  Lahore, 54000, Pakistan.}
\affiliation{Research Center of Astrophysics and Cosmology, Khazar University, Baku, AZ1096, 41 Mehseti Street, Azerbaijan}

\author{Narakorn Kaewkhao}
\email{naragorn.k@psu.ac.th}
\affiliation{Division of Physical Science, Faculty of Science, Prince of Songkla University, Hatyai 90112, Thailand} 

\author{Ali \"Ovg\"un}
\email{ali.ovgun@emu.edu.tr}
\affiliation{Physics Department, Eastern Mediterranean University, Famagusta, 99628 North
Cyprus via Mersin 10, Turkiye.}

\date{\today}

\begin{abstract}

In this study, we investigate traversable wormholes within the framework of Einstein-Euler-Heisenberg (EEH) nonlinear electrodynamics. By employing the Einstein field equations with quantum corrections from the Euler-Heisenberg Lagrangian, we derive wormhole solutions and examine their geometric, physical, and gravitational properties. Two redshift function models are analyzed: one with a constant redshift function and another with a radial-dependent function $\Phi=r_{0}/r$. Our analysis demonstrates that the inclusion of quantum corrections significantly influences the wormhole geometry, particularly by mitigating the need for exotic matter. The shape function and energy density are derived and examined in both models, revealing that the energy conditions, including the weak and null energy conditions (WEC and NEC), are generally violated at the wormhole throat. However, satisfaction of the strong energy condition (SEC) is observed, consistent with the nature of traversable wormholes. The Arnowitt-Deser-Misner (ADM) mass of the EEH wormhole is calculated, showing contributions from geometric, electromagnetic, and quantum corrections. The mass decreases with the Euler-Heisenberg correction parameter, indicating that quantum effects contribute significantly to the wormhole mass. Furthermore, we investigate gravitational lensing within the EEH wormhole geometry using the Gauss-Bonnet theorem, revealing that the deflection angle is influenced by both the electric charge and the nonlinear parameter. The nonlinear electrodynamic corrections enhance the gravitational lensing effect, particularly at smaller impact parameters.

\end{abstract}

\keywords{Shadow Cast, Photon Spheres/Orbits, Static and Rotating Traversable Wormholes}

\maketitle

\newpage
\section{Introduction}
The notion of wormholes—hypothetical tunnels through spacetime linking distant regions of the universe—has captured the imagination of scientists and the public alike since its inception nearly a century ago. First introduced by Einstein and Rosen in 1935 \cite{Einstein:1935tc} as a theoretical consequence of general relativity, these structures, dubbed Einstein-Rosen bridges, hinted at the possibility of shortcuts through space and time, potentially facilitating rapid interstellar travel or even time travel. However, these early wormholes were not traversable; their throats collapsed too swiftly for any matter or information to pass through. A pivotal advancement came in 1988 when Morris and Thorne proposed the concept of traversable wormholes, structures that could remain stable and open for traversal \cite{MT}. Their groundbreaking work revealed a critical requirement: the presence of exotic matter, characterized by negative energy density that violates the weak energy condition (WEC) \cite{Morris:1988tu}. Since exotic matter remains a speculative entity, unobserved in nature, its necessity presents a formidable challenge to the physical plausibility of traversable wormholes \cite{Damour:2007ap, Bueno:2017hyj, Lobo:2005yv, Lemos:2003jb, Sushkov:2005kj, Frolov:1990si, Delgaty:1994vp, Perry:1991qq, Oliva:2009ip, Clement:1983fe, Clement:1995ya, Clement:1997yp, Guendelman:1991pc, Guendelman:2009er, Guendelman:2009pf, Guendelman:2008zp, Tsukamoto:2016qro, Tsukamoto:2016zdu, Tsukamoto:2017hva, Kuhfittig:2013hva, Goulart:2017iko, Ono:2018ybw,Arellano:2008xw, Halilsoy:2013iza, Bronnikov:2017sgg, Shaikh:2018yku, Rahaman:2010kdn, Menchon:2017qed, Harko:2013aya, Bronnikov:2022bud, Arellano:2006np, Sharif:2012zis, Flores-Alfonso:2020nnd, Visser1995, Jusufi:2020rpw, Lobo:2005us, Shaikh:2018kfv, Turimov:2022iff, Volkel:2022khh, Peng:2021osd,Jusufi:2017mav,Rizwan:2024dpp, AraujoFilho:2024iox, Panyasiripan:2024kyu, Samart:2021tvl, Garattini:2023wgk, Javed:2022fsn, Jawad:2022hlm, Javed:2022dfn, Kumaran:2021rgj, Mustafa:2021vqz, Ovgun:2020yuv, Javed:2019qyg, Ovgun:2018prw, Ovgun:2018fnk, Ovgun:2018xys,Yuennan:2024ibh}.

To circumvent this obstacle, physicists have turned to alternative theoretical frameworks that might either reduce the dependence on exotic matter or eliminate it entirely. One such promising approach is nonlinear electrodynamics (NLED), which extends Maxwell’s linear theory by incorporating nonlinear interactions of electromagnetic fields \cite{Sorokin:2021tge}. NLED becomes particularly significant in regimes where quantum effects are pronounced, such as in strong electromagnetic fields or extreme gravitational environments. Among NLED models, the Euler-Heisenberg (EH) theory stands out \cite{Heisenberg:1936nmg,Yajima:2000kw}, rooted in quantum electrodynamics (QED). The EH Lagrangian accounts for quantum corrections arising from virtual particle pairs, offering a framework to explore how these effects influence gravitational phenomena in strong-field contexts.

In this paper, we explore traversable wormholes within the framework of Einstein-Euler-Heisenberg (EEH) nonlinear electrodynamics. By coupling the EH Lagrangian to the Einstein field equations, we derive wormhole solutions and investigate their geometric, physical, and gravitational properties. Our study considers two distinct models for the redshift function: one featuring a constant redshift and another with a radial-dependent form, $\Phi = r_0/r$, where $r_0$ is a characteristic length scale. Through meticulous analysis, we demonstrate that the quantum corrections embedded in the EH theory profoundly impact the wormhole geometry, notably by mitigating the reliance on exotic matter. We derive the shape function and energy density for both models, providing insights into the spatial structure and matter distribution of these wormholes. Our examination of energy conditions reveals that the weak energy condition (WEC) and null energy condition (NEC) are generally violated at the wormhole throat—a hallmark of traversable wormholes—while the strong energy condition (SEC) is partially satisfied, consistent with theoretical expectations.

Beyond geometry and energy considerations, we compute the Arnowitt-Deser-Misner (ADM) mass of the EEH wormhole, uncovering contributions from its geometry, electromagnetic fields, and quantum corrections. Notably, the ADM mass increases with the Euler-Heisenberg correction parameter, highlighting the significant role of quantum effects in the wormhole’s total mass. Additionally, we investigate the gravitational lensing properties of these wormholes using the Gauss-Bonnet theorem. Our findings indicate that the deflection angle of light rays traversing near the wormhole depends intricately on both the electric charge and the nonlinear parameter, with the nonlinear electrodynamic corrections amplifying the lensing effect, particularly at smaller impact parameters. Understanding these lensing signatures could offer a pathway to observationally distinguish wormholes from other compact objects, such as black holes, should they exist in the universe.

This paper is organized as follows: Section 2: Presents the theoretical foundation, detailing the EEH action and the general metric for static, spherically symmetric wormholes. Section 3: Derives the wormhole solutions for the two redshift function models. Section 4: Analyzes the energy conditions and their implications for the matter content sustaining the wormhole. Section 5: Calculates the ADM mass and explores its dependence on model parameters. Section 6: Examines the gravitational lensing effects, emphasizing the influence of nonlinear electrodynamics. Section 7: Summarizes our findings and suggests directions for future research. Through this study, we aim to deepen the understanding of traversable wormholes in a quantum-corrected gravitational framework, shedding light on their feasibility and observable signatures.

\section{EEH Traversable Wormholes}
Constructing Einstein-Euler-Heisenberg (EEH) wormhole solutions involves solving Einstein's field equations coupled with the Euler-Heisenberg nonlinear electrodynamics. This formulation extends the usual Einstein-Maxwell theory by incorporating quantum corrections from quantum electrodynamics (QED). The Euler-Heisenberg Lagrangian describes quantum corrections to classical electrodynamics due to vacuum polarization effects in strong electromagnetic fields \cite{Hamil:2024njs}. The Euler-Heisenberg Lagrangian includes quantum corrections given as \cite{Yajima:2000kw}
\begin{eqnarray}
\mathcal{L}_{\text{EH}} &=& -\frac{1}{4} F_{\mu\nu} F^{\mu\nu} + \alpha (F_{\mu\nu} F^{\mu\nu})^2 + \beta (F_{\mu\nu} \tilde{F}^{\mu\nu})^2\nonumber\\&=&-\frac{1}{4}F+\alpha F^{2}+\beta G^{2}\,.\label{EH}
\end{eqnarray}
The first term features the classical Maxwell Lagrangian, with corresponding energy-momentum tensor:
\begin{eqnarray}
T^{\rm EM}_{\mu\nu}=F_{\mu\alpha}F^{\alpha}_{\nu}-\frac{1}{4}g_{\mu\nu}F_{\gamma\delta} F^{\gamma\delta}\,.\label{EM}
\end{eqnarray}
Here, the arguments of ${\cal L}_{\text{EH}}$ are the electromagnetic field invariants
\begin{eqnarray}
F\equiv F_{\mu\nu} F^{\mu\nu}\,,G\equiv F_{\mu\nu} \tilde{F}^{\mu\nu}\,,
\end{eqnarray}
where $F_{\mu\nu}$ is the electromagnetic field tensor, ${\tilde F}_{\mu\nu}$ is its dual and $\alpha,\beta$ are QED one-loop correction coefficients. The action of EEH non-linear electrodynamics including a quantum correction to the linearized Maxwell’s theory is written as
\ba
{\cal S}_{\rm EEH}= \int d^{4}x\sqrt{-g}\left(\frac{R}{16\pi} + \mathcal{L}_{\text{EH}}\right)\,,
\ea
where $R$ stands for the Ricci scalar, $G=1$ (Gravitational constant) and $\mathcal{L}_{\text{EH}}$ is given in Eq.(\ref{EH}). The Einstein field equations in the presence of Euler-Heisenberg nonlinear electrodynamics are given by
\begin{equation}
G_{\mu\nu} = 8\pi T_{\mu\nu}=8\pi (T^{\rm EM}_{\mu\nu}+T^{\rm QC}_{\mu\nu})\,,
\end{equation}
where $G_{\mu\nu}$ is the Einstein tensor representing spacetime curvature, $T_{\mu\nu}$ is the total stress-energy tensor, and $T^{\rm QC}_{\mu\nu}$ denotes the stress-energy tensor govern by the loop (quantum) corrections. For a spherically symmetric metric, we use the standard wormhole form
\begin{equation}
ds^2 = -e^{2\Phi(r)} dt^2 + \frac{dr^2}{1 - \frac{b(r)}{r}} + r^2 d\Omega^2\,,
\end{equation}
where in this context, $\Phi(r)$ and $b(r)$ represent the redshift and shape functions, respectively. To ensure the absence of an event horizon within the wormhole geometry, the redshift function $\Phi(r)$ must remain finite. Meanwhile, the shape function $b(r)$ characterizes the wormhole's geometry and must satisfy the condition $b(r_0) = r_0$, where $r_0$ denotes the radius of the wormhole throat. Additionally, the shape function must fulfill the flaring-out condition \cite{MT,Morris:1988tu}:
\ba
\frac{b(r) - r b'(r)}{b^2(r)} > 0\,,
\ea
where the derivative of the shape function, $b'(r) = \frac{db}{dr}$, must satisfy $b'(r) < 1$ at the wormhole throat, and  $d\Omega^{2}=d\theta^{2}+\sin^{2}\theta d\phi^{2}$ is the $2$-sphere metric. Substituting the chosen form of 
$\Phi(r)$ and $b(r)$, and the Lagrangian into the Einstein equations, we obtain
\begin{widetext}
\begin{eqnarray}
\frac{b'(r)}{r^2} &=& 8\pi \rho(r)\,,\\
-\frac{b(r)}{r^3} + 2 \left( 1 - \frac{b(r)}{r} \right) \frac{\Phi'(r)}{r} &=& 8\pi p_r(r)\,,\label{pr}\\
\left(1 - \frac{b}{r} \right) \left[\Phi'' + (\Phi')^2 + \frac{\Phi'}{r} - \frac{b'r - b}{2r(r-b)} \right] &=& 8\pi p_t(r)\,.\label{pt}
\end{eqnarray}
\end{widetext}
To compute the stress-energy tensor in Euler-Heisenberg nonlinear electrodynamics, we use:
\begin{eqnarray}
T_{\mu\nu} &=& g_{\mu\nu} \mathcal{L}_{\text{EH}} - 4 \frac{\partial \mathcal{L}_{\text{EH}}}{\partial F} F_{\mu\tau} F_{\nu}^{\ \tau} \nonumber\\&-& 4 \frac{\partial \mathcal{L}_{\text{EH}}}{\partial G} \tilde{F}_{\mu\tau} F_{\nu}^{\ \tau}\,.
\end{eqnarray}
The total energy density $\rho_{\rm EH}$ in the Einstein-Euler-Heisenberg (EEH) wormhole is derived from the stress energy tensor associated with Euler-Heisenberg non-linear electrodynamics (NLED). This density arises from both the Maxwellian and quantum-corrected terms in the Euler-Heisenberg Lagrangian. For a static, spherically symmetric wormhole, the energy density is obtained from the $(0,0)$ component
\begin{equation}
-T^{0}_{0}\equiv\rho = \frac{1}{2} ({\bf E}^2 + {\bf B}^2) + 2 \alpha ({\bf E}^2 - {\bf B}^2)^2 + 2 \beta ({\bf E} \cdot {\bf B})^2\,,\label{rho}
\end{equation}
where $E=F_{tr}$ is the electric field,
$B={\tilde F}_{tr}$	is the magnetic field, and $\alpha ({\bf E}^2 - {\bf B}^2)^2$ represents nonlinear quantum corrections. Notice that the energy density $\rho$ contains the standard electric and magnetic energy densities including corrections from quantum effects via $\alpha ({\bf E}^2 - {\bf B}^2)^2$, which can modify the exotic matter requirement. The radial pressure $p_{r}$ is typically negative for wormhole solutions, indicating the necessity of exotic matter. The quantum correction term $\alpha ({\bf E}^2 - {\bf B}^2)^2$ may help balance the equation, reducing the need for exotic matter. Additionally, the radial pressure $p_{r}$ and transverse pressure $p_{t}$ are also influenced by the nonlinear EH terms arising from EH electrodynamics. Note that The $\beta$-term would contribute if both $E$ and $B$ exist simultaneously, and
they are not perpendicular, so that ${\bf E}\cdot {\bf B}\neq 0$.

\subsection{Model with $\Phi={\rm constant}$}
The simplest case is a model with $\Phi={\rm constant}$, namely a spacetime with no tidal forces, namely $\Phi'(r) = 0$. This choice is commonly adopted in the literature as it streamlines the analysis while preserving physically significant solutions. By setting $\Phi=0$, gravitational redshift effects are completely eliminated, which not only simplifies the metric but also the associated field equations. More importantly, this assumption guarantees the absence of event horizons, thereby maintaining the wormhole's traversability. Numerous studies on traversable wormholes have utilized this approach; see Refs.\cite{MT,Morris:1988tu,Visser1995}. For a pure electric charge $E(r)=q/r^{2}\,,B=0$, we find
\begin{eqnarray}
b(r)=c_{1}+\frac{32 \pi  \alpha  q^4}{5 r^5}+\frac{q^2}{r}\,.
\end{eqnarray}
Finally we use $b(r_{0}) = b_{0} = r_{0}$, to calculate the the
constant $c_{1}$. Thus by solving Eq.(\ref{br}), we find the shape function to be
\begin{eqnarray}
b(r)&=&r_{0}+q^2 \left(\frac{1}{r}-\frac{1}{r_{0}}\right)\nonumber\\&+&
\frac{32 \pi  \alpha  q^4}{5}\left(\frac{1}{r^5}-\frac{1}{r^5_{0}}\right)\,.\label{br}
\end{eqnarray}

\begin{figure}
\includegraphics[width=8 cm]{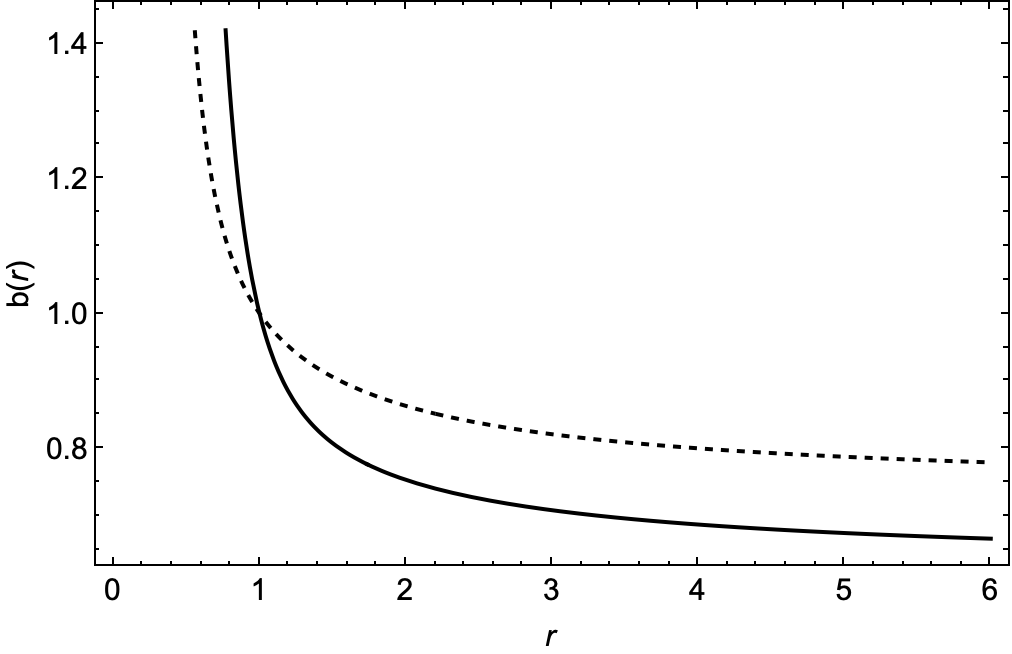}
\caption{The shape function of the EEH wormhole against $r$. Here we have used $r_0=1$, and $\alpha=0.01$ (dashed line) and $\alpha=0.1$ (solid line).}\label{fig1}
\end{figure}
In Fig.\ref{fig1}, we have plotted the shape function $b(r)$ against $r$. Introducing the scaling of coordinate $\exp(2\Phi)dt^{2} \rightarrow dt^{2}$ and taking $\Phi={\rm constant}$, the wormhole metric reads
\begin{eqnarray}
ds^2 &=& -dt^2 \notag\\ &+& \frac{1}{1 - \frac{r_{0}}{r}+\frac{q^2}{r} \left(\frac{1}{r_{0}}-\frac{1}{r}\right)+
\frac{32 \pi  \alpha  q^4}{5r}\left(\frac{1}{r_{0}^5}-\frac{1}{r^5}\right)}dr^2 \notag\\&+& r^2 d\Omega^2\,. \label{whsol1}
\end{eqnarray}
Clearly in the limit $r\rightarrow \infty$, we obtain
\begin{eqnarray}
\lim_{r\rightarrow \infty}\frac{b(r)}{r}=0\,.
\end{eqnarray}
\begin{figure}
\includegraphics[width=8 cm]{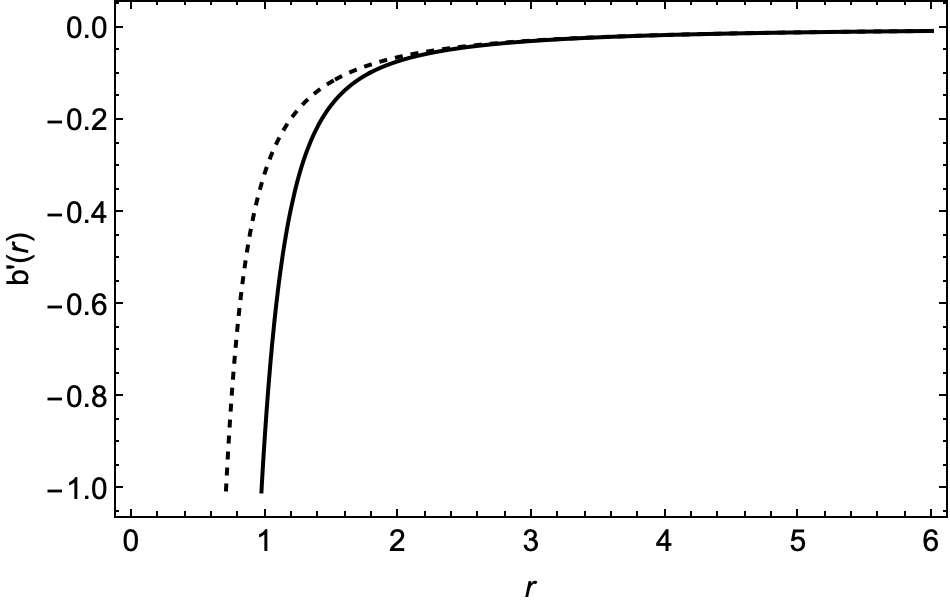}
\caption{We check the flare out condition and variation of $b'(r)$ against $r$ is displayed. Here we have used $r_0=1$, and $\alpha=0.01$ (dashed line) and $\alpha=0.1$ (solid line).  }\label{fig2}
\end{figure}
The asymptotically flat metric can be seen also from Fig.\ref{fig2}. Using the EoS $p_r(r)=\omega(r) \rho(r)$, one can easily see that when $\Phi(r)=0$ (tideless wormholes) we obtain 
\begin{equation}\label{24}
8 \omega(r) \rho(r) \pi r^3+b(r)=0.
\end{equation}
Solving this equation  for the EoS parameter, we obtain
\begin{figure}
\includegraphics[width=8 cm]{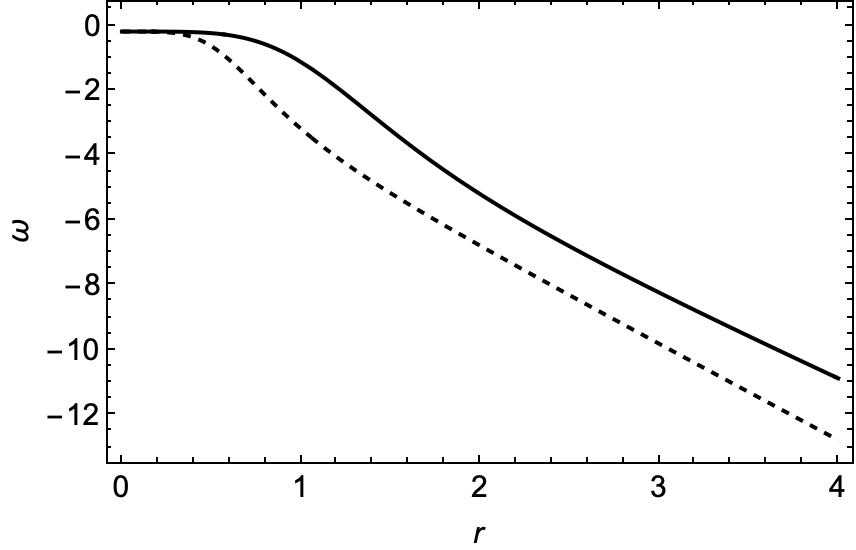}
\caption{The EoS parameter $\omega$ for the EEH wormhole with $\Phi=0$ as a function of $r$. Here we have used $r_0=1$, and $\alpha=0.01$ (dashed line) and $\alpha=0.1$ (solid line).}\label{ome}
\end{figure}

\begin{equation}
    \omega=\frac{32 \pi  \alpha  q^4 \left(r_{0}^5-r^5\right)-5 r^4 r_{0}^4 \left(q^2 (r-r_{0})+r r_{0}^2\right)}{5 q^2 r_{0}^5 \left(8 \pi  \alpha  q^2+r^4\right)}.\label{e27}
\end{equation}
The plot in Fig.\ref{ome} shows the radial dependence of a parameter $\omega$ as a function of the radial coordinate $r$. We notice that for larger values of $r$, the EoS parameter becomes significantly negative, indicating that the radial pressure is dominated by negative values, typical of exotic matter sustaining the wormhole throat.

\subsection{Model with $\Phi=r_{0}/r$}
This form of the redshift function represents a non-trivial gravitational potential that remains finite throughout the spacetime, including at the throat (r=r0). By ensuring that the metric coefficient $g_{tt}$ remains non-vanishing, this choice effectively prevents the formation of event horizons and facilitates smooth traversal through the wormhole. Moreover, this functional form is motivated by previous studies that investigated wormholes supported by various exotic matter distributions and examined their stability under perturbations (see Refs. \cite{Lobo:2005us, Jusufi:2020rpw}). The choice $\Phi=r_{0}/r$ introduces a redshift effect that gradually diminishes at large distances, making it particularly relevant in astrophysical contexts where the gravitational potential naturally decays with radial distance.
 
\subsubsection{EoS: $p_r(r)=\omega_r(r) \rho(r)$}
We shall begin our analysis by considering the following EoS $p_r(r)=\omega_r(r) \rho(r)$. From the Einstein's field equations (\ref{pr}), we find
\begin{equation}\label{24}
\Phi'(r)=-\frac{8 \omega_r(r) \rho(r) \pi r^3+b(r)}{2 r(-r+b(r))}.
\end{equation}
Now considering the model function
\begin{equation}
    \Phi(r)=\frac{r_0}{r},\label{e26}
\end{equation}
we obtain the following equation
\begin{eqnarray}\notag
\frac{(r-2 r_0)b(r)+8 \omega_r(r) \rho(r) r^4 \pi+2 r_0 r}{8 \pi r^4}=0.
\end{eqnarray}
Finally using the shape function (\ref{br}) for the EoS parameter we obtain
\ba
\omega_r(r)&=&-\frac{5 q^2 r^4 (2 r_{0}-1) r_{0}^4 (r-r_{0})}{5 q^2 r_{0}^5 \left(32 \pi  \alpha  q^2+r^4\right)}\notag\\&+&\frac{5 r^5 (2 r_{0}-3) r_0^6}{5 q^2 r_{0}^5 \left(32 \pi  \alpha  q^2+r^4\right)}\notag\\&+&\frac{32 \pi  \alpha  q^4 (2 r_{0}-1) \left(r_{0}^5-r^5\right)}{5 q^2 r_{0}^5 \left(32 \pi  \alpha  q^2+r^4\right)}\,.\label{e27}
\ea

\begin{figure}
\includegraphics[width=8 cm]{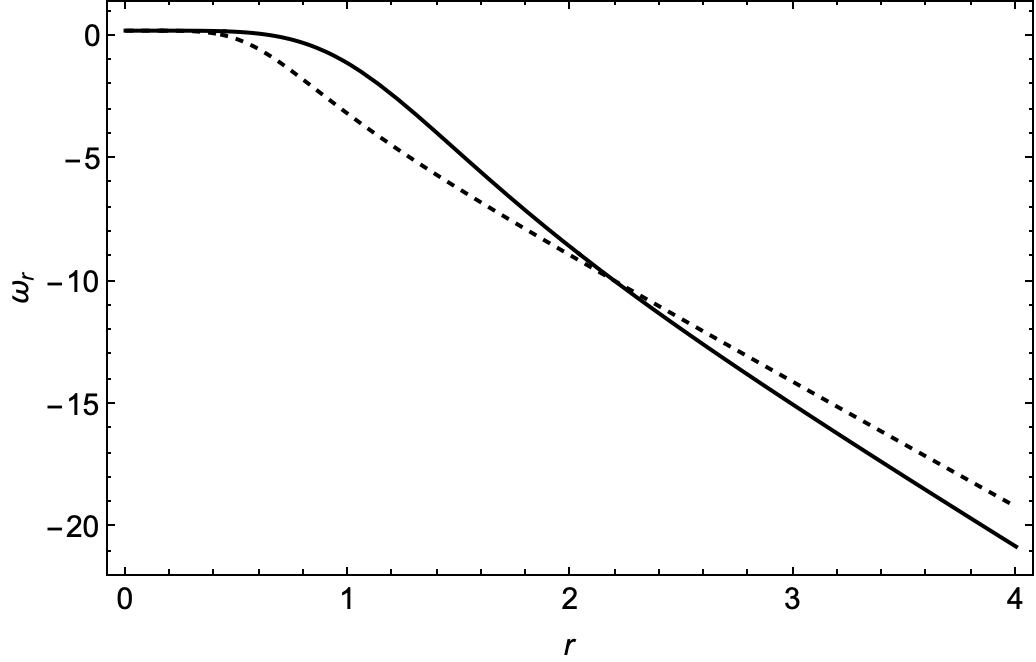}
\caption{The EoS parameter $\omega_r(r)$ against $r$ with a non-constant redshift
function $\Phi=r_0/r$ as a function of $r$. Here we have used $r_0=1$, and $\alpha=0.01$ (dashed line) and $\alpha=0.1$ (solid line) along with a non-constant redshift
function $\Phi=r_0/r$. }\label{fig4}
\end{figure}

The behavior of the EoS parameter $\omega_r(r)$ for the EEH wormhole with a non-constant redshift function $\Phi=r_0/r$ as a function of $r$ is displayed in Fig.\ref{fig4}.

\subsubsection{EoS: $p_t(r)=\omega_t (r) p_r(r)$}
Let us now consider the scenario in which the EoS is of the form $p_t(r)=\omega_t(r) p_r(r)$, where $\omega_t(r)$ is as an arbitrary function of $r$ . In this case, combining the second (\ref{pr}) and the third equation (\ref{pt}), we find the following relation
\begin{eqnarray}\notag
   &2& r(r+1)(r-b(r))\Phi''(r)+2r^2 (r-b(r))(\Phi'(r))^2\\\notag
   &-&r \Phi'(r)\left[ (-4 \omega_t(r)+r-1)b(r)+4 \omega_t(r) r \right]\\
   &+&b(r)(2 \omega_t(r) -r +1)=0.
\end{eqnarray}

\begin{figure}
\includegraphics[width=8 cm]{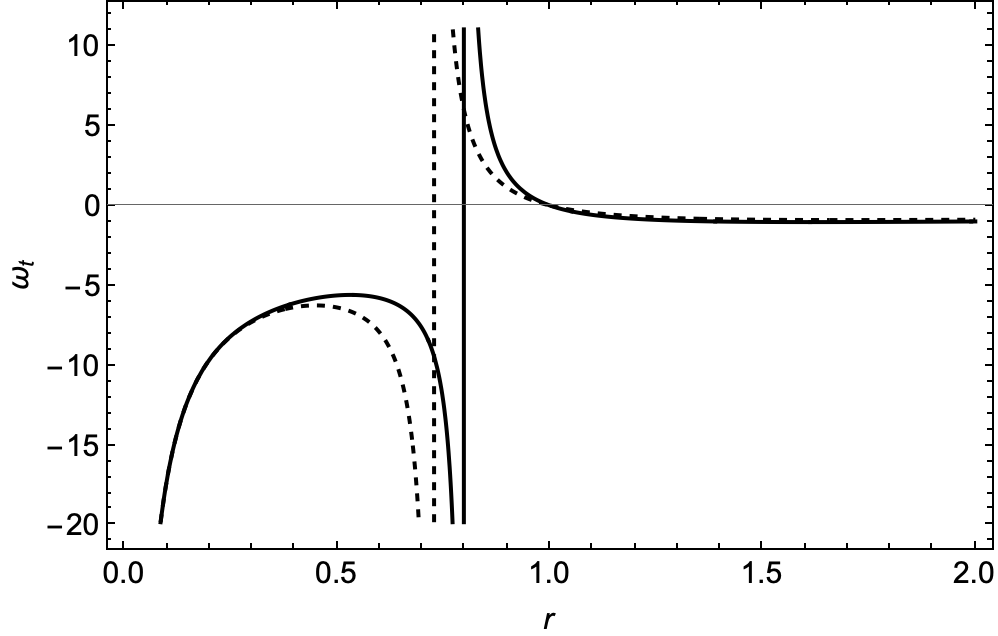}
\caption{The EoS parameter $\omega_t(r)$ for the EEH wormhole with a non-constant redshift
function $\Phi=r_0/r$ as a function of $r$. Here we have used $r_0=1$, and $\alpha=0.01$ (dashed line) and $\alpha=0.1$ (solid line) along with a non-constant redshift
function $\Phi=r_0/r$.  }\label{fig5}
\end{figure}
Using the shape function (\ref{br}) along with Eq. (\ref{e26}) from the last equation we obtain 
\begin{equation}
   \omega_t(r)= \frac{(r-r_{0}) \left(32 \pi  \alpha  A B q^4+5 r^4 r_{0}^4 S\right)}{10 C r^5 r_{0}^4+64 \pi  \alpha  H q^4 r},\label{e30}
\end{equation}
where
\begin{eqnarray}
  A &=&r^3-r^2 (r_{0}+1)+5 r r_{0}+2 r_{0} (r_{0}+2),\notag\\B &=&r^4+r^3 r_{0}+r^2 r_{0}^2+r r_{0}^3+ r_{0}^4,\notag\\
    C &=& q^2 (r-r_{0}) (r-2 r_{0})+r r_{0}^2 (3 r-2 r_{0}),\notag\\H &=& r^6-2 r^5 r_{0}-r r_{0}^5+2 r_{0}^6\,,\notag\\S&=&\left(A q^2+r r_{0}^2 ((r-5) r-2 (r_{0}+2))\right)\,.\notag
\end{eqnarray}
It is straightforward to check the dependence of $\omega_{r}(r)$ and $\omega_{t}(r)$ against $r$ given by Fig.\ref{fig4} and Fig.\ref{fig5}, respectively.

\subsection{Isotropic model with $\omega_r(r)={\rm const.}$}

From the conservation equation \(\nabla_{\mu}T^{\mu\nu}=0\), we derive the hydrostatic equilibrium equation for the matter sustaining the wormhole as follows:
\begin{eqnarray}
    p'_r(r) &=& \frac{2(p_t(r) - p_r(r))}{r} \nonumber\\&-& (\rho(r) + p_r(r))\Phi'(r),
\end{eqnarray}
where we have assumed a perfect fluid with equal radial and tangential pressures (\(p_t = p_r\)) and adopt the equation of state (EoS) given by \(p_r(r) = \omega_r \rho(r)\), where \(\omega_r\) is a constant parameter. This equation simplifies to:
\begin{equation}
    \omega_r \rho'(r) = -(1 + \omega_r) \rho(r) \, \Phi'(r),
\end{equation}
in which \(\rho(r)\) is specified by Eq. (\ref{rho}). Solving this differential equation, we obtain:
\begin{equation}
    \Phi(r) = C + \frac{\omega_r}{\omega_r + 1} \ln\left(\frac{r^8}{r^4 + 32 \pi \alpha q^2}\right).
    \label{iso}
\end{equation}
By absorbing the constant \(C\) via the scaling \(dt \to C \, dt\), the wormhole metric element can be expressed as:
\begin{eqnarray} 
\mathrm{d}s^{2} &=& -\left(\frac{r^8}{r^4 + 32 \pi \alpha q^2}\right)^{\frac{2}{1 + 1/\omega_r}} \mathrm{d}t^{2} + \frac{\mathrm{d}r^{2}}{1 - \frac{b(r)}{r}}\nonumber\\&+& r^{2} \left( \mathrm{d}\theta^{2} + \sin^{2}\theta \, \mathrm{d}\phi^{2} \right),  
\end{eqnarray}
where \(r \geq r_0\). It is straightforward to observe that the solution remains finite at the wormhole throat (\(r = r_0\)) provided $\omega_r \neq -1$. Nevertheless, the redshift function \(\Phi(r)\) becomes unbounded for large \(r\), implying that asymptotically flat EEH wormholes with isotropic pressures cannot be constructed, and such solutions may generally lack physical viability.

\subsection{Anisotropic model with $\omega_r=const.$}

The isotropic model, as previously demonstrated, holds minimal physical significance. In this final example, we construct an anisotropic Einstein-Euler-Heisenberg (EEH) wormhole spacetime. To achieve this, we introduce the following relationships for the pressures:
\[
p_t(r) = n \, \omega_r \, \rho(r), \quad p_r(r) = \omega_r \, \rho(r),
\]
where \(n\) is a constant parameter. This leads to the following differential equation:
\begin{eqnarray}
    \omega_r \, \rho'(r) &=& \frac{2 \, \omega_r \, \rho(r) \, (n - 1)}{r} \nonumber\\&-& (1 + \omega_r) \, \rho(r) \, \Phi'(r).
\end{eqnarray}
By solving this equation for the redshift function \(\Phi(r)\), we obtain
\begin{equation}
    \Phi(r) = C + \frac{\omega_r}{\omega_r + 1} \ln\left(\frac{r^{2 (n + 3)}}{r^4 + 32 \pi \alpha q^2}\right).
\end{equation}
The resulting metric takes the form:
\begin{eqnarray}
\mathrm{d}s^{2} &=& -\left(\frac{r^{2 (n + 3)}}{r^4 + 32 \pi \alpha q^2}\right)^{\frac{2}{1 + 1/\omega_r}} \mathrm{d}t^{2} + \frac{\mathrm{d}r^{2}}{1 - \frac{b(r)}{r}} \nonumber\\&+& r^{2} \left(\mathrm{d}\theta^{2} + \sin^{2}\theta \, \mathrm{d}\phi^{2}\right),  
\end{eqnarray}
where the radial coordinate satisfies \(r \geq r_0\). The isotropic case described earlier Eq.(\ref{iso}) is recovered by setting \(n = 1\), while a singularity emerges for \(\omega_r = -1\). In the anisotropic scenario, it can be shown that asymptotically flat spacetime configurations are possible.

\begin{figure}
\includegraphics[width=8 cm]{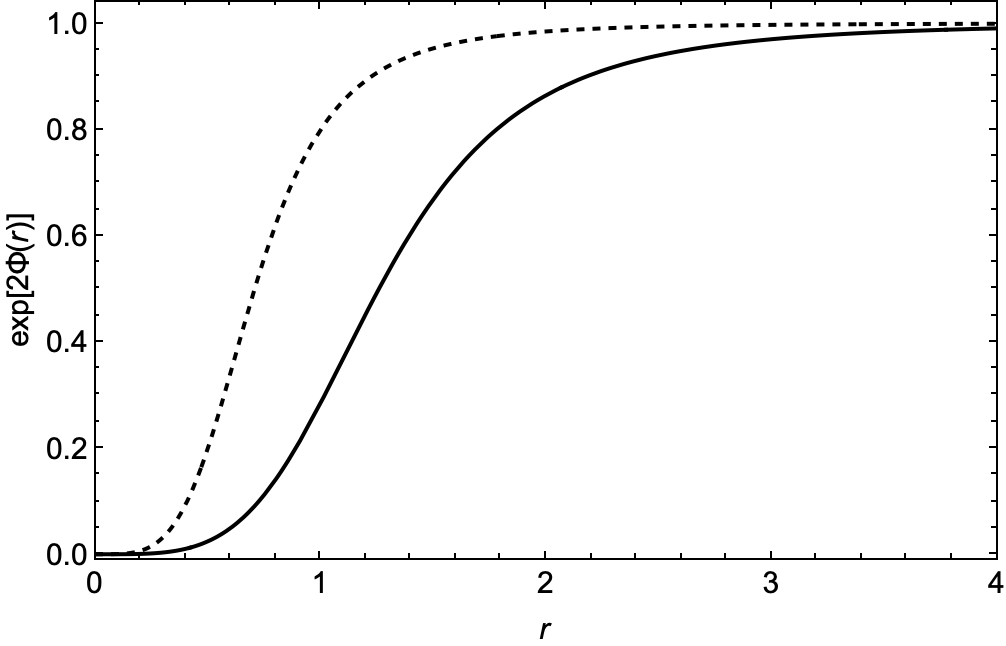}
\caption{We plot $\exp(2\Phi(r))$ for the anisotropic case as a function of $r$. Here we have used $r_0=1,\,q=0.5$, and $\alpha=0.01$ (dashed line) and $\alpha=0.1$ (solid line).}\label{f88}
\end{figure}

By setting \(n = -1\) and ensuring \(\omega_r \neq -1\), the metric simplifies to:
\begin{eqnarray}
\mathrm{d}s^{2} &=& -\left(\frac{1}{1 + \frac{32 \pi \alpha q^2}{r^4}}\right)^{\frac{2}{1 + 1/\omega_r}} \mathrm{d}t^{2} + \frac{\mathrm{d}r^{2}}{1 - \frac{b(r)}{r}} \nonumber\\&+& r^{2} \left(\mathrm{d}\theta^{2} + \sin^{2}\theta \, \mathrm{d}\phi^{2}\right),  
\end{eqnarray}
This metric describes an asymptotically flat spacetime, and in fact, the choice of \(n = -1\) uniquely leads to an asymptotically flat solution. As illustrated in Fig.\ref{f88}, the limiting behavior as \(r \to \infty\) results in \(\exp(2\Phi(r)) = 1\), consistent with the expected outcome.

\section{Embedding Diagram}
\label{Embed}
\begin{figure}
\includegraphics[width=8.0 cm]{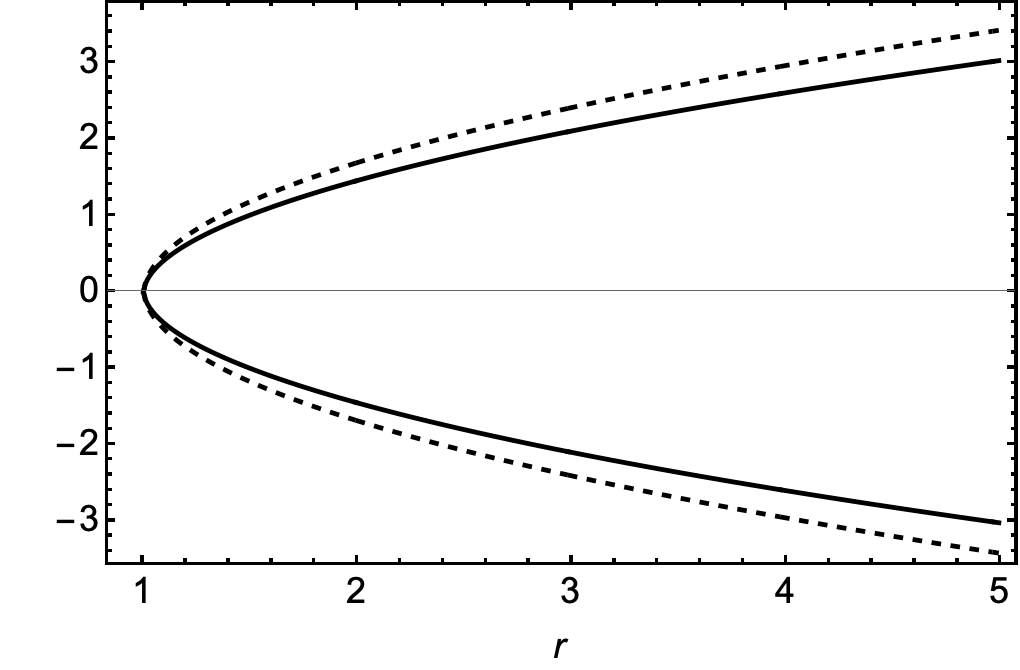}
\caption{The 2D embedding diagram of the EEH wormhole. Here we have used $r_0=1,\,q=0.5$, and $\alpha=0.01$ (dashed line) and $\alpha=0.1$ (solid line).}\label{f8}
\end{figure}

In this section, we explore the embedding diagrams used to depict the EEH wormhole by examining an equatorial slice at $\theta=\pi/2$ for a fixed moment in time, where $t=\text{constant}$. The metric of this slice can be expressed as:
\begin{equation}
    ds^2 = \frac{dr^2}{1 - \frac{b(r)}{r}} + r^2 d\phi^2. \label{emb}
\end{equation}
To visualize this spatial slice, we embed the given metric (\ref{emb}) into a three-dimensional Euclidean space. The resulting spacetime can be formulated in cylindrical coordinates as:
\begin{equation}
    ds^2 = dz^2 + dr^2 + r^2 d\phi^2.
\end{equation}
By comparing the above two equations, we obtain the following relation:
\begin{equation}
    \frac{dz}{dr} = \pm \sqrt{\frac{r}{r - b(r)} - 1}.
\end{equation}
Here, $b(r)$ is determined by Eq.(\ref{br}). It is important to note that the integral of this expression cannot be performed analytically. Therefore, numerical methods are used to illustrate the geometry of the wormhole as shown in Fig.\ref{f8}. The figure reveals how the quantum correction parameter $\alpha$ influences the geometry of the wormhole.

\section{ADM Mass of EEH wormhole}
\label{ADM}
\begin{figure}
\includegraphics[width=8.0 cm]{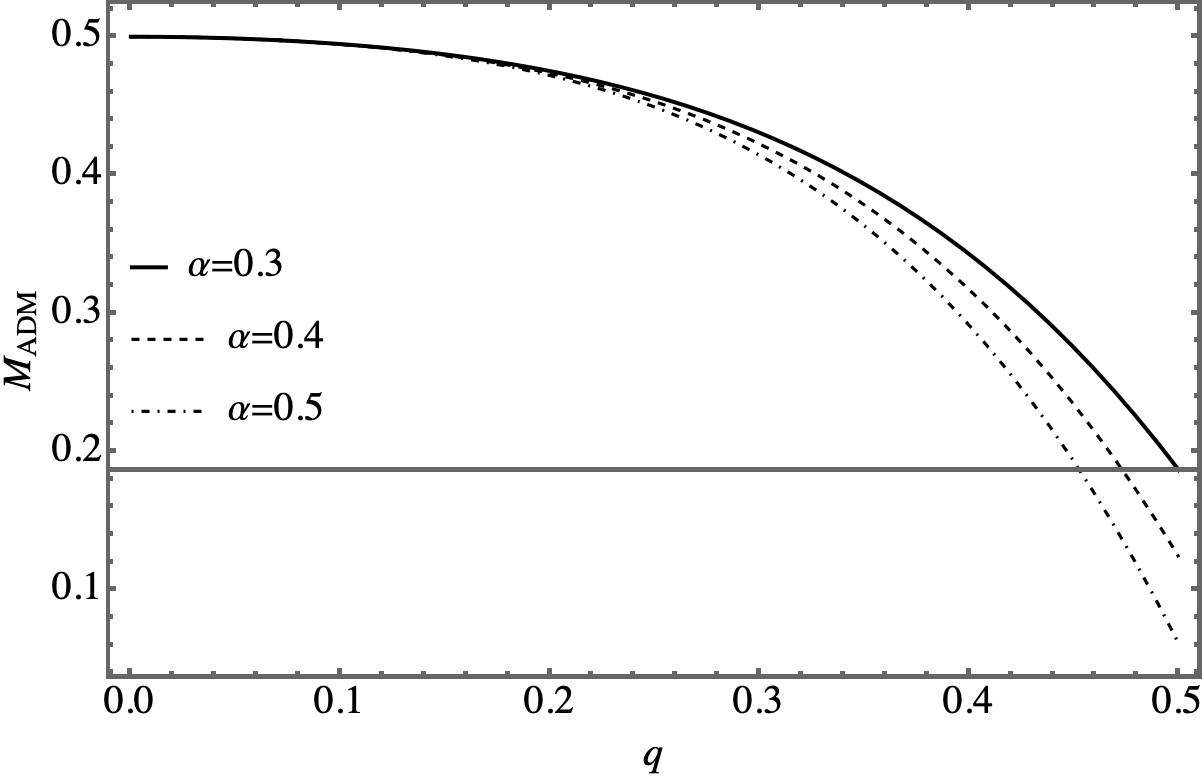}
\caption{The ADM mass $M_{\rm ADM}$ is plotted as a function of the black hole charge $q$ for three different values of the model parameter $\alpha$: 0.3 (solid line), 0.4 (dashed line), and 0.5 (dash-dotted line). The figure illustrates that $M_{\rm ADM}$ decreases monotonically with increasing charge for all values of $\alpha$, indicating a reduction in the total mass-energy of the system as the electric charge grows.}\label{fm}
\end{figure}

Let us now calculate the Arnowitt-Deser-Misner (ADM) mass for the EEH wormhole. We consider an asymptotically flat spacetime given by the metric
\begin{eqnarray}
ds^2_{\Sigma} &=& \psi(r) \, dr^2 \nonumber\\&+& r^2 \chi(r) \left( d\theta^2 + \sin^2\theta \, d\phi^2 \right),
\end{eqnarray}
where the metric functions are defined as
\begin{equation}
    \psi(r) = \frac{1}{1 - \frac{b(r)}{r}}, \quad \text{and} \quad \chi(r) = 1.
\end{equation}
To compute the ADM mass, we employ the following relation (see \cite{Shaikh:2018kfv}):
\begin{equation}\label{for}
    M_{\rm ADM} = \lim_{r \to \infty} \frac{1}{2} \left[-r^2 \chi' + r (\psi - \chi) \right].
\end{equation}
By substituting the respective values into the above formula and evaluating the limit, we obtain the ADM mass of the wormhole as
\begin{equation}\label{ADM}
    M_{\rm ADM} = \frac{r_{0}}{2} - \frac{q^2}{2 r_{0}} - \frac{16 \pi \alpha q^4}{5 r_{0}^5}.
\end{equation}
This represents the mass of the wormhole as perceived by an observer located at asymptotic spatial infinity. It is evident that the quantum correction parameter $\alpha$ decreases the ADM mass. Notably, the ADM mass (\ref{ADM}) consists of three distinct contributions: the first term corresponds to the geometric component ($r_0$), the second term arises from the charge effect, and the third term captures the quantum effects within the spacetime. The ADM mass is plotted Fig.(\ref{fm}) as function of charge $q$ with different values of $\alpha$. The results show the simultaneous effect of charge and the quantum corrections.

The inclusion of the Euler-Heisenberg correction parameter $\alpha$ has a direct and significant impact on the ADM (Arnowitt-Deser-Misner) mass of the wormhole. In the Einstein-Euler-Heisenberg (EEH) framework, the ADM mass incorporates contributions from geometric, electromagnetic, and quantum (nonlinear electrodynamic) effects. As $\alpha$ increases, the third term becomes more negative, thereby reducing the total ADM mass of the wormhole. This decrease in ADM mass implies that quantum effects via EH corrections contribute a repulsive or stabilizing influence, offsetting the energy needed to sustain the wormhole. It suggests that quantum vacuum fluctuations encoded in nonlinear electrodynamics can play a structural role, partially replacing exotic matter and enhancing the stability and physical plausibility of the wormhole.

\section{Energy Conditions}
\label{Ener}
Given the redshift function and the shape function, we can determine the components of the energy-momentum tensor. Specifically, the radial component is calculated for $\Phi=r_{0}/r$ as follows:
\ba
p_r &=& \frac{q^2 (r-r_{0}) (r-2 r_{0})+r r_{0}^2 (2 r_{0}-3 r)}{8 \pi  r^5 r_{0}}\notag\\&+&\frac{4 \alpha  q^4 \left(r^6-2 r^5 r_{0}-r r_{0}^5+2 r_{0}^6\right)}{5 r^9 r_{0}^5}\,.
\ea
In contrast, the tangential pressure component is given by:
\ba
p_t &=& \frac{1}{80 \pi  r^{10} r_{0}^5}\Big(5 r^4 r_{0}^4 \big(-q^2 \big(r^4-2 (r+1) r^2 r_{0}+2 r_{0}^3\big)\notag\\&+&r^3 (r+2) r_{0}^2-2 r r_{0}^4\big)-32 \pi  \alpha  q^4 \big(r^8-2 r^6 r_{0}-2 r^5 r_{0}^2\notag\\&-&6 r^3 r_{0}^5+2 r r_{0}^6+2 r_{0}^7\big)\Big)\,.
\ea
\begin{figure}
\includegraphics[width=8cm]{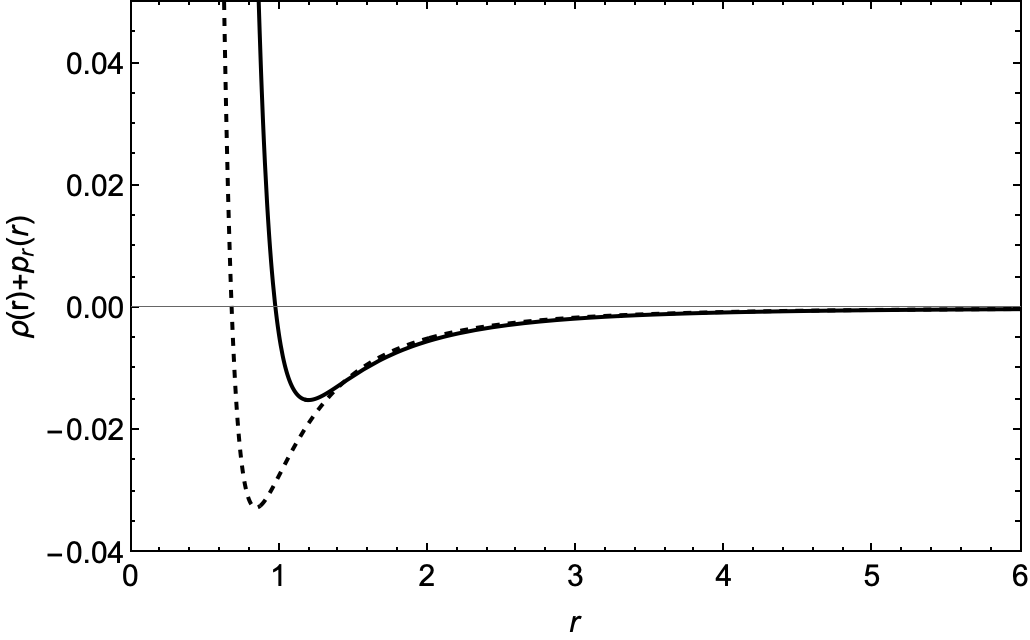}
\caption{The variation of \(\rho + p_r\) as a function of \(r\) using \(\Phi = r_0 / r\). Here we have used $r_0=1,\,q=0.5$, and $\alpha=0.01$ (dashed line) and $\alpha=0.1$ (solid line).}\label{wec}
\end{figure}
\begin{figure}
\includegraphics[width=8cm]{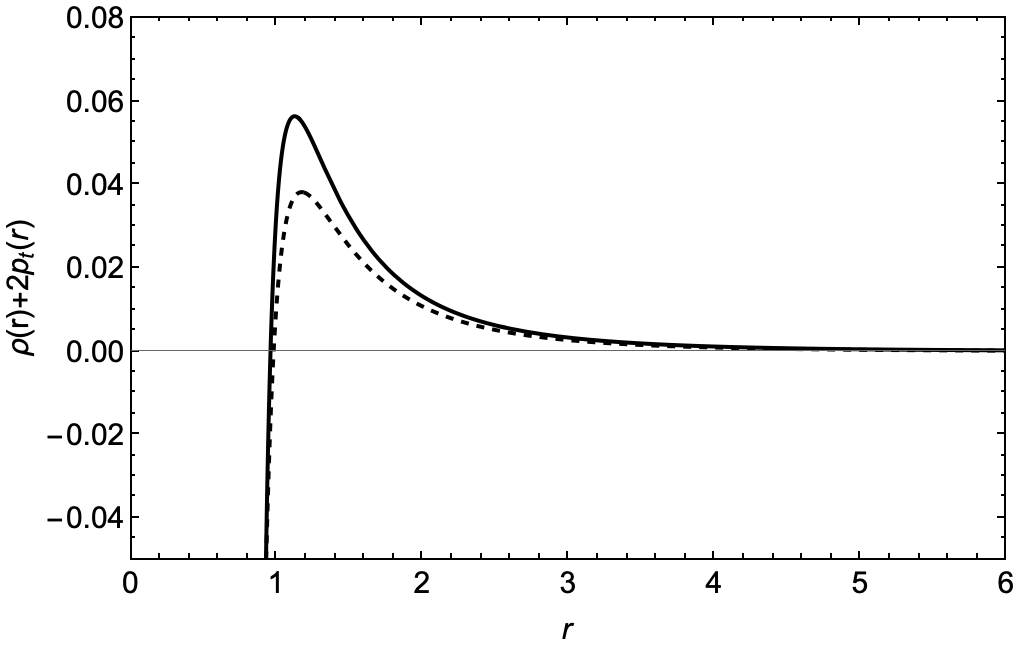}
\caption{The variation of \(\rho + 2p_t\) against \(r\) and \(\Phi = r_0 / r\). Here we have used $r_0=1,\,q=0.5$, and $\alpha=0.01$ (dashed line) and $\alpha=0.1$ (solid line).}
\label{fig6}\label{sec}
\end{figure}
\begin{figure}
\includegraphics[width=8cm]{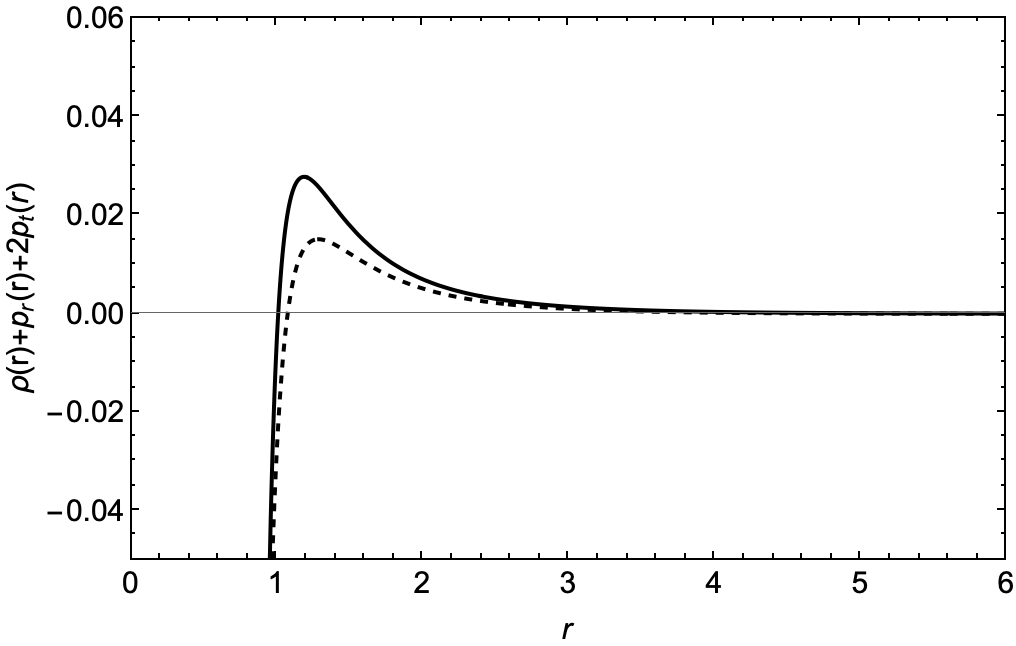}
\caption{The variation of \(\rho + p_r + 2 p_t\) against \(r\) and \(\Phi = r_0 / r\). Here we have used $r_0=1,\,q=0.5$, and $\alpha=0.01$ (dashed line) and $\alpha=0.1$ (solid line).}
\label{sec1}
\end{figure}
With these results, we proceed to analyze the energy conditions and evaluate their validity through regional plots. The weak energy condition (WEC) is characterized by:
\[
T_{\mu \nu }U^{\mu }U^{\nu } \geq 0,
\]
which implies:
\[
\rho(r) + p_r(r) \geq 0,
\]
where \(T_{\mu \nu }\) denotes the energy-momentum tensor and \(U^{\mu }\) represents a timelike vector. This indicates that the local energy density is nonnegative, and the null energy condition (NEC) defined as $T_{\mu \nu }k^{\mu }k^{\nu } \geq 0$, where \(k^{\mu }\) is a null vector:
\[
\rho(r) + p_r(r) \geq 0.
\]
Furthermore, the strong energy condition (SEC) requires:
\[
\rho(r) + 2p_t(r) \geq 0,
\]
and
\[
\rho(r) + p_r(r) + 2p_t(r) \geq 0.
\]
From the figures, it is evident that the WEC and NEC are not satisfied at the wormhole throat (\(r = r_0\)), see Fig.\ref{wec}, while the SEC are satisfied at the wormhole throat (\(r = r_0\)), Fig.\ref{sec} \& \ref{sec1}. Additionally, its value strongly depends on the $\alpha$ parameter. Numerical evaluations reveal that at the throat (\(r_0 = 1\)), we have:
\[
(\rho + p_r)\vert_{r_0=1} < 0,
\]
and
\[
(\rho + p_r + 2p_t)\vert_{r_0=1} > 0,
\]
with only minor deviations. In the context of quantum field theory, it is well-established that quantum fluctuations often violate conventional energy conditions without any constraints, suggesting that such fluctuations could play a vital role in the stability of wormholes. One can ask how do quantum corrections from the Euler-Heisenberg Lagrangian affect the violation or satisfaction of classical energy conditions in wormhole geometries? In Einstein-Euler-Heisenberg (EEH) gravity, quantum corrections arise through the nonlinear electrodynamics (NLED) terms of the Euler-Heisenberg Lagrangian. These corrections, characterized primarily by the parameter $\alpha$ in the term $\alpha ({\bf E}^2 - {\bf B}^2)^2$, significantly influence the stress-energy tensor supporting the wormhole geometry. Additionally, the EH corrections represent one-loop QED effects, introducing vacuum polarization and nonlinear electromagnetic interactions. These quantum effects act as a source of modified pressure and energy density, which reshape the energy distribution supporting the wormhole. As a result, the EH framework provides a more physically plausible model, where the role of exotic matter is partially replaced by quantum vacuum effects.

We also observed that the choice of redshift function—whether constant or radially dependent—plays a critical role in shaping the geometry of the wormhole and in determining the extent to which exotic matter is required to sustain the structure. In the context of Einstein-Euler-Heisenberg (EEH) gravity, this choice directly affects the metric components, the behavior of tidal forces, and the distribution of energy density and pressures derived from the modified field equations. A constant redshift function ($\Phi=\rm{const.}$) simplifies the geometry and ensures traversability by eliminating tidal forces. However, it leads to strong violations of the weak and null energy conditions (WEC and NEC), indicating a higher dependence on exotic matter. In contrast, a radial-dependent redshift function (e.g., $\Phi=r_{0}/r$) introduces spatial variation in the gravitational potential, which enhances the flexibility of the spacetime geometry. This choice allows quantum corrections from the Euler-Heisenberg Lagrangian to more effectively balance the energy-momentum components, leading to partial satisfaction of the strong energy condition (SEC) and weaker violations of WEC and NEC. As a result, the need for exotic matter is significantly reduced.

\section{Light deflection by EEH wormhole}
\label{def}
In this subsection, we investigate weak gravitational lensing within the EEH wormhole geometry using the Gauss-Bonnet theorem (GBT) \cite{Gibbons:2008rj}. To achieve this, we first consider null geodesics ($\mathrm{d}s^{2}=0$) constrained to the equatorial plane ($\theta=\pi/2$), leading us to define an optical metric suitable for this analysis: \begin{equation} \mathrm{d}t^{2}=\frac{1}{A(r)^2}\mathrm{d}r^2+\frac{C(r)}{A(r)}\mathrm{d}\varphi^2. \end{equation}

We then introduce a new radial coordinate, known as the tortoise coordinate $r^{\star}$, defined by the transformation \begin{equation} \mathrm{d}r^{\star }=\frac{1}{A(r)}\mathrm{d}r, \end{equation} and a function $f(r^{\star})$ related to the radial dependence of the optical metric, given by \begin{equation} f^2(r^{\star })=\frac{C(r)}{A(r)}. \end{equation}

Using these new definitions, we rewrite the EEH wormhole optical metric compactly as: \begin{equation} \mathrm{d}t^{2}=\tilde{g}_{ab},\mathrm{d}x^{a}\mathrm{d}x^{b}=\mathrm{d}{r^{\star }}^{2}+f^{2}(r^{\star })\mathrm{d}\varphi ^{2}, \quad (a,b=r,\varphi). \end{equation}

Next, we calculate the Gaussian optical curvature $\mathcal{K}$, crucial for applying the GBT. Explicitly, the Gaussian curvature can be expressed in terms of the function $f(r^{\star})$ as follows: 
\begin{eqnarray} \mathcal{K} & = & - \frac{1}{f (r^{\star})} \frac{\mathrm{d}^2 f(r^{\star})}{\mathrm{d} r^{\star 2}} \\ \notag & = & - \frac{1}{f (r^{\star})} \left[ \frac{\mathrm{d} r}{\mathrm{d} r^{\star}} \frac{\mathrm{d}}{\mathrm{d} r} \left( \frac{\mathrm{d} r}{\mathrm{d} r^{\star}} \right) \frac{\mathrm{d} f}{\mathrm{d} r} + \left( \frac{\mathrm{d} r}{\mathrm{d} r^{\star}} \right)^2 \frac{\mathrm{d}^2 f}{\mathrm{d} r^2} \right]. \end{eqnarray}

\subsection{Case with $\Phi(r)={\rm const.}$}

To analyze gravitational lensing in the weak deflection regime, we adopt a weak-field approximation on the optical metric of \ref{whsol1}, allowing us to simplify the Gaussian curvature into a manageable form:
\ba\label{Curvature1} 
\mathcal{K} =  -\frac{96 \pi  \alpha  q^4}{5 r^8}+\frac{16 \pi  \alpha  q^4}{5 r^3 r_0^5}-\frac{q^2}{r^4}+\frac{q^2}{2 r^3 r_0}-\frac{r_0}{2 r^3}
\ea
With the curvature approximation at hand, we now proceed to find the deflection angle by applying the Gauss-Bonnet theorem. For this purpose, we consider a non-singular region $\mathcal{D}{R}$ bounded by the curve $\partial\mathcal{D}{R}=\gamma {\tilde{g}}\cup C{R}$. This domain permits us to apply the GBT, which states \cite{Gibbons:2008rj}: \begin{equation} \iint\limits_{\mathcal{D}{R}}\mathcal{K},\mathrm{d}S+\oint\limits{\partial \mathcal{ D}{R}}\kappa ,\mathrm{d}t+\sum{i}\theta {i}=2\pi \chi (\mathcal{D}{R}), \end{equation} where $\kappa$ is the geodesic curvature, $\mathcal{K}$ the Gaussian curvature, and $\theta_{i}$ are exterior angles at vertices. By construction, we choose $\mathcal{D}{R}$ to have an Euler characteristic number $\chi (\mathcal{D}{R})=1$ (simply connected domain).

To explicitly find the deflection angle, it is necessary to compute the geodesic curvature $\kappa$. We use the definition of geodesic curvature along the curve $\gamma$, given by \begin{equation} \kappa =\tilde{g}\left(\nabla {\dot{\gamma}}\dot{\gamma},\ddot{\gamma}\right), \end{equation} where $\tilde{g}(\dot{\gamma},\dot{\gamma})=1$ is the unit-speed condition, and $\ddot{\gamma}$ is the acceleration vector. Considering the asymptotic limit ($R\rightarrow\infty$), the angles at the source $\mathcal{S}$ and observer $\mathcal{O}$ approach right angles, thus their sum $\theta_{\mathcal{O}}+\theta_{\mathcal{S}}\rightarrow\pi$. In this limit, the GBT simplifies to: \begin{equation} \iint\limits_{\mathcal{D}{\infty}}\mathcal{K},\mathrm{d}S+\oint\limits{C_{\infty}}\kappa,\mathrm{d}t=\pi-\int\limits_{0}^{\pi+\Theta}\mathrm{d}\varphi. \end{equation}

We now explicitly evaluate the geodesic curvature $\kappa(C_{R})$ on the curve $C_{R}=r(\varphi)=R=\text{constant}$. Considering its definition, we have: \begin{equation} \kappa (C_{R})=|\nabla {\dot{C}{R}}\dot{C}_{R}|. \end{equation}

Computing the radial component of this curvature explicitly yields: \begin{equation} \left( \nabla {\dot{C}{R}}\dot{C}{R}\right)^{r}=\dot{C}{R}^{\varphi },\partial {\varphi }\dot{C}{R}^{r}+\tilde{\Gamma} {\varphi\varphi}^{r}\left( \dot{C}{R}^{\varphi}\right)^{2}. \end{equation}

Clearly, the first derivative term vanishes as $r(\varphi)=\text{constant}$. Using the explicit form of $\tilde{\Gamma}{\varphi\varphi}^{r}$ and the unit-speed condition, we find in the asymptotic limit: \begin{eqnarray}\notag \lim{R\rightarrow \infty }\kappa (C_{R}) =\lim_{R\rightarrow \infty }\left|\nabla {\dot{C}{R}}\dot{C}_{R}\right|\approx\frac{1}{R}. \end{eqnarray}%

Moreover, at large radial distances, we approximate the arc-length element along $C_{R}$ by: \begin{eqnarray} \lim_{R\rightarrow \infty}\mathrm{d}t &\approx& R\mathrm{d}\varphi. \end{eqnarray}%

Combining these asymptotic expressions gives a simple result for the integral along the boundary curve: \begin{equation} \kappa(C_{R})\mathrm{d}t\approx\mathrm{d}\varphi. \end{equation}

For convenience in evaluation, we choose the deflection line described by the relation $r(\varphi)=b/\sin\varphi$, where $b$ denotes the impact parameter. With this parameterization, the expression for the deflection angle $\Theta$ using the GBT finally reduces to the integral form \cite{Gibbons:2008rj}: \begin{eqnarray}\label{int0} \Theta=-\int\limits_{0}^{\pi}\int\limits_{\frac{b}{\sin\varphi}}^{\infty}\mathcal{K}\mathrm{d}S. \end{eqnarray}

By explicitly inserting the approximate Gaussian curvature from Eq. \eqref{Curvature1} into this integral, and applying the simplification $dr^{\star}\approx dr$ valid at large distances, we can evaluate the integral up to the second-order terms, yielding the final deflection angle expression: 

\begin{equation}\label{GB10} \Theta \approx -\frac{32 \pi  \alpha  q^4}{5 b r_0^5}-\frac{q^2}{b r_0}+\frac{r_0}{b}. \end{equation}

\begin{figure}
\includegraphics[width=8 cm]{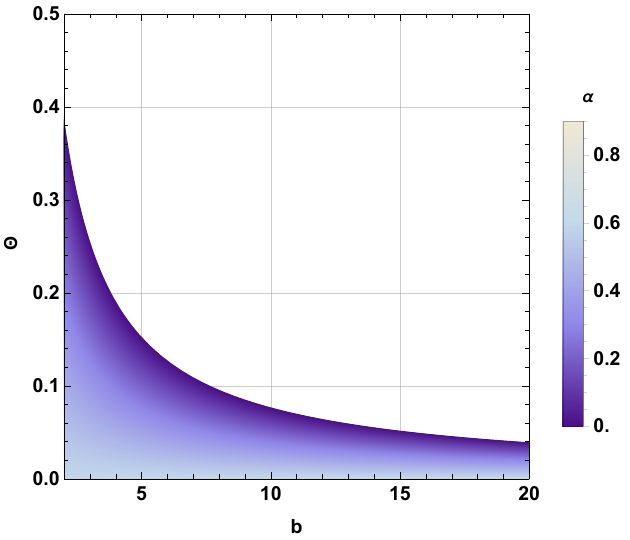}
\caption{Figure shows the deflection angle versus impact parameter for the EEH wormhole case with $\Phi(r)={\rm const.}$ ($r_0=1,\,q=0.5$) }\label{wdawh1}
\end{figure}

This result encapsulates various physical influences arising from geometric and electromagnetic contributions, including nonlinear electrodynamic effects attributed to the Euler-Heisenberg coupling parameter, $\alpha$ shown in Fig. \ref{wdawh1}. Each term in this expression can be interpreted distinctly, revealing valuable insights into the physical nature of such wormholes. 

Here, the term proportional to \(\alpha q^4\) encapsulates the nonlinear electromagnetic corrections, entering with a negative sign and thus contributing a defocusing effect. In contrast, the conventional charge term \(-q^2/(b r_0)\) similarly diminishes the deflection angle relative to the purely geometric contribution, \(\frac{r_0}{b}\). This balance illustrates that, in the absence of additional gravitational redshift effects, the lensing properties are governed by a competition between the geometric curvature induced by the wormhole throat and the modifications introduced by both the standard and nonlinear electromagnetic fields.

\subsection{Isotropic model with $\omega_r(r)={\rm const.}$}

In the isotropic model where the radial equation-of-state parameter, \(\omega_r\), is constant, the deflection angle takes the form

\begin{equation}\label{GB20} \Theta \approx \frac{128 \pi  \alpha  q^4 \omega_r }{5 b r_0^5}-\frac{32 \pi  \alpha  q^4}{5 b r_0^5}+\frac{4 q^2 \omega_r }{b r_0}-\frac{q^2}{b r_0}-\frac{4 r_0 \omega_r }{b}+\frac{r_0}{b} . \end{equation}

\begin{figure}
\includegraphics[width=8 cm]{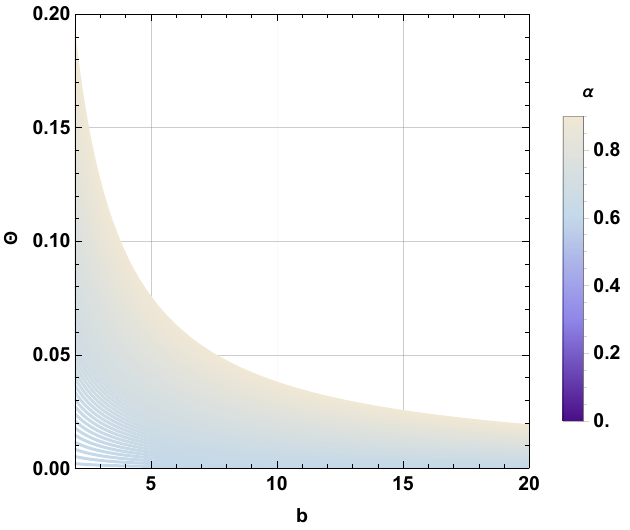}
\caption{Figure shows the deflection angle versus impact parameter for the EEH wormhole case with Isotropic model with $\omega_r(r)={\rm const.}$ ($r_0=1,\,q=\omega_r=0.5$) }\label{wdawh2}
\end{figure}

The additional terms proportional to \(\omega_r\) reflect the presence of isotropic matter within the wormhole throat and decrease the deflection angle as shown in Fig. \ref{wdawh2}. Notably, the nonlinear electromagnetic correction is now modulated by \(\omega_r\) in the first term, suggesting that the matter content can enhance the deflection angle when \(\omega_r\) is positive. Moreover, the appearance of modified charge contributions and a corresponding geometric correction further indicates that the internal structure of the wormhole has a nontrivial impact on the lensing properties. This result underscores the importance of matter anisotropy (or, in this case, its absence) in shaping the observable features of wormhole spacetimes.

\subsection{Anisotropic model with $\omega_r=const.$}

Lastly, we calculate the deflection angle in weak field limits for the anisotropic model as follows:

\begin{eqnarray}\label{GB30} \Theta \approx  \frac{9 \pi ^2 \alpha  q^4}{32 b^4 r_0^2}+\frac{3 \pi  q^4}{512 b^4}-\frac{48 \pi ^2 \alpha  q^2 \omega_r }{b^4}+\frac{16 \pi  \alpha  q^4}{15 b^3 r_0^3}+\frac{q^4}{12 b^3 r_0} \notag \\-\frac{q^2 r_0}{6 b^3}+\frac{4 \pi ^2 \alpha  q^4}{5 b^2 r_0^4}-\frac{\pi  q^4}{16 b^2 r_0^2}-\frac{\pi  q^2}{8 b^2}+\frac{32 \pi  \alpha  q^4}{5 b r_0^5}+\frac{q^2}{b r_0}-\frac{r_0}{b}. \end{eqnarray}

\begin{figure}
\includegraphics[width=8 cm]{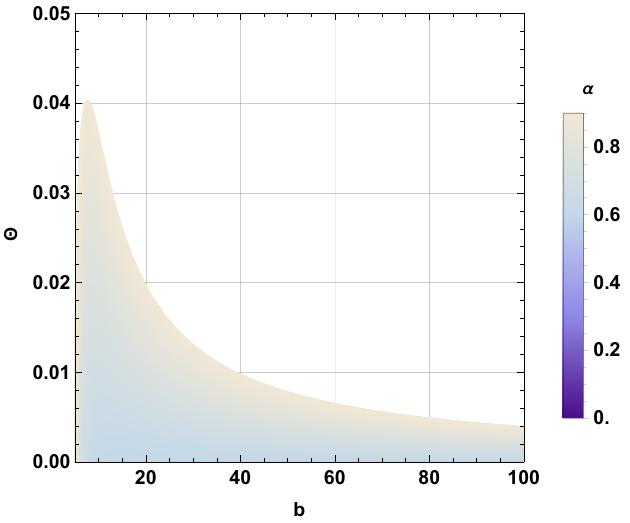}
\caption{Figure shows the deflection angle versus impact parameter for the EEH wormhole case with anisotropic model with $\omega_r=const.$ ($r_0=1,\,q=\omega_r=0.5$) }\label{wdawh3}
\end{figure}

The proliferation of terms in this expression reflects the interplay of higher-order corrections and the effects of anisotropy in the stress-energy tensor. In particular, cross-terms combining \(\alpha\) and \(\omega_r\) indicate that the anisotropic pressure distribution not only modifies the electromagnetic contributions but also alters the effective geometric curvature experienced by light rays as shown in Fig. \ref{wdawh3} that make smaller the deflection angle than other models. The dependence on various inverse powers of \(b\) and \(r_0\) implies that the deflection angle is sensitive to the scale of the impact parameter, with higher-order corrections becoming relevant in different regimes. Such richness in structure suggests that observational signatures of anisotropic wormholes could be markedly distinct from their isotropic or constant redshift counterparts.

In summary, our analysis highlights that in the weak field regime the deflection angle for traversable wormholes in Einstein-Euler-Heisenberg nonlinear electrodynamics is highly sensitive to both the geometric configuration and the properties of the matter threading the wormhole. The contrasting roles of the nonlinear electrodynamics coupling, the electromagnetic charge, and the matter pressure parameter \(\omega_r\) offer a multifaceted phenomenology.

Analyzing these contributions collectively, it becomes apparent that the deflection angle’s behavior significantly depends on the intricate balance between quantum, electromagnetic, and geometric factors. Quantum electrodynamics introduces nuanced effects predominantly at small impact parameters, while classical geometric contributions dominate at larger distances.

Consequently, observing and measuring such gravitational lensing effects could provide valuable empirical insights into distinguishing EEH wormholes from other astrophysical compact objects. Additionally, since the Euler-Heisenberg parameter $\alpha$ emerges from nonlinear electrodynamics and quantum corrections, detecting such deviations in observed deflection angles could open a new window into testing quantum gravity scenarios and nonlinear electromagnetic theories experimentally.

In the Euler–Heisenberg framework, nonlinear electrodynamics modifies gravitational lensing by altering the Gaussian curvature of the optical metric through vacuum‐polarization corrections: in the weak‐field expansion, terms proportional to $\alpha\,q^4/r^8$ and $\alpha\,q^4/(r^3r_0^5)$ arise directly from the quartic invariants in the EH Lagrangian and introduce higher‐order inverse–power dependencies that are absent in the linear Maxwell case.  When these corrections are inserted into the Gauss–Bonnet theorem (GBT) integral, they yield an extra defocusing contribution $-32\pi\alpha\,q^4/(5\,b\,r_0^5)$ to the total deflection angle, in addition to the standard Maxwell term $-q^2/(b\,r_0)$ and the purely geometric wormhole term $r_0/b$.  Geometrically, the GBT cleanly separates the EH‐induced curvature from the asymptotically trivial boundary term—since the nonlinear corrections vanish at infinity, they affect only the interior curvature integral—thereby isolating the signature of vacuum polarization in the lensing profile and offering a clear pathway to distinguish nonlinear electromagnetic effects from classical geometry and linear charge contributions.

\section{Conclusion}
In this work, we examined the geometry and physical properties of traversable wormholes within the framework of Einstein--Euler--Heisenberg (EEH) nonlinear electrodynamics. We analyzed wormhole solutions under two redshift function models---a constant and a radially varying form $\Phi = r_0/r$---focusing on energy conditions, ADM mass, embedding structure, and gravitational lensing phenomena.

A central finding is that quantum corrections arising from the Euler--Heisenberg Lagrangian significantly alter wormhole geometry and stability. The nonlinear coupling parameter $\alpha$ modifies the energy density and pressures in a way that reduces the exotic matter requirement. While the weak and null energy conditions (WEC and NEC) remain violated at the throat, the strong energy condition (SEC) is partially satisfied, aligning with theoretical expectations for traversable wormholes. The ADM mass, composed of geometric, electromagnetic, and quantum contributions, exhibits a marked dependence on both charge and $\alpha$, showing that quantum corrections play a crucial role in defining the total energy of the system. Similarly, gravitational lensing analysis reveals that the deflection angle is sensitive to these corrections, especially at smaller impact parameters, providing potential observational signatures.

Our results suggest that nonlinear electrodynamics, particularly through EEH corrections, can alleviate some of the exotic matter constraints and introduce detectable features in lensing and shadow profiles. These findings support the viability of wormholes as astrophysical candidates under extended gravity frameworks. Future work should explore dynamic stability and time-dependent solutions, along with potential observational signals arising from particle motion and wave perturbations.

Further directions include analyzing test particle trajectories near EEH wormholes~\cite{Turimov:2022iff,Tangphati:2023mpk} and investigating their quasi-normal mode spectra~\cite{Volkel:2022khh}, which could reveal unique signatures distinct from black holes. Studies of optical observables such as lensing rings and brightness distributions~\cite{Peng:2021osd} may also help differentiate asymmetric thin-shell wormholes from compact objects. Additionally, recent intersecting works~\cite{Ditta:2024iky,Ditta:2024lnb,Ashraf:2024xti,Mustafa:2024kjy,Mustafa:2024mvx,Ashraf:2024xwm,Ditta:2025vsa,Ashraf:2024itj,Ahmed:2024jsh,Ahmed:2024qeu,Ahmed:2025oaw} emphasize the broader role of quantum corrections and modified gravity in shaping the behavior of black holes and wormholes through geodesic motion, thermodynamics, and stability analyses.

Looking ahead, advancing observational techniques---such as those employed by the Event Horizon Telescope (EHT)~\cite{EventHorizonTelescope:2019ggy} and the forthcoming Legacy Survey of Space and Time (LSST)\footnote{\url{http://www.lsst.org/lsst}}---could offer critical tests of these theories. Distinctive features in wormhole shadows, deflection patterns, polarization, and gravitational wave signals may serve as powerful tools to distinguish wormholes from black holes. Future efforts should aim to refine these theoretical predictions and establish connections with measurable astrophysical phenomena, especially in scenarios involving rotating wormholes, anisotropic fluids, and noncommutative or higher-dimensional gravitational frameworks.

\acknowledgements
A.{\"O}. would like to acknowledge the contribution of the COST Action CA21106 - COSMIC WISPers in the Dark Universe: Theory, astrophysics and experiments (CosmicWISPers), the COST Action CA21136 - Addressing observational tensions in cosmology with systematics and fundamental physics (CosmoVerse), the COST Action CA22113 - Fundamental challenges in theoretical physics (THEORY-CHALLENGES), the COST Action CA23130 - Bridging high and low energies in search of quantum gravity (BridgeQG) and the COST Action CA23115 - Relativistic Quantum Information (RQI) funded by COST (European Cooperation in Science and
Technology), and A.{\"O}. would like to thank EMU, TUBITAK, Turkiye and SCOAP3, Switzerland for their support.

\end{document}